\documentclass[12pt]{article}

\usepackage{graphicx}

\begin{document}

\title{\Large \bf Low mass diffractive systems at LHC }
\author{\large R. Schicker$^1$, \bigskip \\
{\it  $^1$~Phys.Inst., Philosophenweg 12, 69120 Heidelberg} }

\maketitle


\begin{center}
{\bf Abstract}\\
\medskip

Diffractive reactions in proton-proton  collisions are characterized by the
presence of rapidity gaps and by forward scattered protons.
A diffractive trigger can therefore be defined by the tagging of the 
forward proton or by the detection of rapidity gaps.
I present a diffractive trigger scheme for the ALICE detector
at the large hadron collider LHC and discuss some physics topics.
In particular, I concentrate on the low mass sector in central
exclusive diffraction which becomes accessible by a double gap trigger.   

\end{center}

\section{The ALICE detector} 
\label{sec:ALICE}

The ALICE experiment at the LHC is designed as a general purpose 
experiment  with a central barrel covering the pseudorapidity range 
$ -0.9 < \eta < 0.9$ and a muon spectrometer covering the range 
$ -4.0 < \eta < -2.5$ \cite{Alice1,Alice2}. 
The ALICE experimental program foresees data taking in pp and PbPb 
collisions
at luminosities of $\mathcal L$ = 5x10$^{30}$cm$^{-2}$s$^{-1}$ and
$\mathcal L$ = 10$^{27}$cm$^{-2}$s$^{-1}$, respectively. 
An asymmetric system pPb will be measured at a luminosity of 
$\mathcal L$ = $10^{29}$cm$^{-2}$s$^{-1}$.

The central detectors track and identify particles from $\sim$ 
100 MeVc$^{-1}$  to  $\sim$ 100 GeVc$^{-1}$ transverse momenta. 
Short-lived particles such as hyperons, D and B mesons are 
identified by their reconstructed secondary decay vertex. The detector 
granularity is chosen such that these tasks can be performed in a high 
multiplicity environment of up to 8000 charged particles per unit of 
rapidity. Tracking of particles is achieved by the inner tracking 
system (ITS) of two layers of silicon pixel (SPD), two layers of
silicon strip (SSD) and two layers of silicon drift detectors (SDD).
The global reconstruction of particle momentum uses the ITS
information together with the information from a large
Time-Projection-Chamber (TPC) and a high granularity 
Transition-Radiation Detector (TRD). 
Particle identification in the central barrel is performed by
measuring energy loss  in the tracking detectors, transition  
radiation in the TRD and  time-of-flight in a high-resolution TOF array. 
A single arm High-Momentum Particle Identification Detector
(HMPID) with limited solid angle coverage extends the momentum 
range of identified hadrons. Photons will be measured by a crystal 
PbWO$_{4}$ PHOton Spectrometer (PHOS) and an electromagnetic sampling
calorimeter (EMCAL). 

\begin{figure}[h]
\centering
\includegraphics[width=8cm,height=6cm,bb = 1 130 596 716,clip]{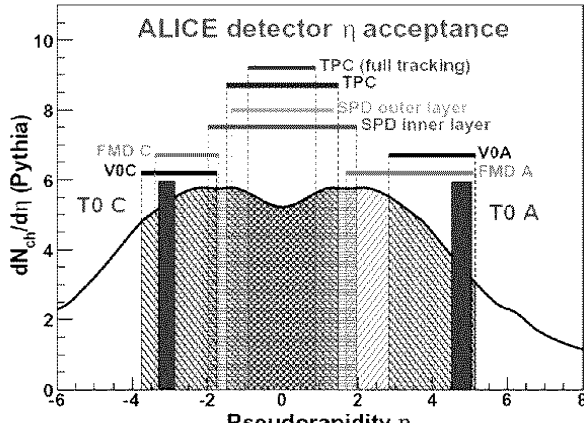} 
\caption{Pseudorapidity coverage of the different detector systems of ALICE}
\label{pict1}
\end{figure}

Additional detectors for trigger purposes and for event classification 
are placed on both sides of the central barrel such that the
pseudorapidity range $ -3.7 < \eta < 5 $ is covered.
Fig. \ref{pict1} shows the pseudorapidity acceptance of ALICE
resulting from the ALICE detectors as explained above. The event 
characterization detectors shown in this figure are quartz
scintillation detectors (T0A,T0C) used for timing, plastic 
scintillator detectors (V0A,V0C) and silicon detectors (FMD) 
for multiplicity characterization.

At very forward angles, a neutron calorimeter is placed on both sides 
of ALICE at a distance of 116 m from the interaction point\cite{ZDC}.
The LHC dipoles which are located between this detector and the
interaction point deflect the charged particles such that only neutral
particles emitted at 0$^0$ reach this detector. 

\subsection{The diffractive gap trigger in ALICE}
\label{sec:gaptrig}

The ALICE trigger is designed as a  system with three levels L0,L1,L2 
and a high-level software trigger (HLT). A diffractive L0 trigger can be 
defined by requiring little or no activity in the V0A and
V0C detectors explained above. These two detectors are designed with
an eight and four-fold  segmentation in azimuth and pseudorapidity,
respectively. The segmentation in pseudorapidity allows to select  
a gap width in steps of half a unit up to the maximum pseudorapidity 
interval of two covered by the detectors. 

The high-level trigger has access to the 
information of all the detectors shown in Fig. \ref{pict1} and hence can 
increase the rapidity gap to the range $-3.7 < \eta <-0.9$ and 
$0.9 < \eta < 5.0$, respectively.

Due to the absence of a V0A and V0C signal in a diffractive trigger,
the L0 signal for this trigger has to be defined within the central
barrel. In defining a L0 diffractive trigger, the transition radiation
detector needs special consideration. This detector system is put in 
sleep mode after readout of an event in order to reduce power
consumption. A wakeup signal is necessary to activate the onboard 
readout electronics. The V0A and V0C signals are transfered to the 
TRD pretrigger system where such a wakeup signal is generated. 

A L0 diffractive trigger can, for example, be defined by the silicon 
pixel detector of the inner tracking system. This signal is, however, 
not in time for the wakeup call of the transition radiation detector. 
A TRD diffractive wakeup call can be defined by the information of the
time-of-flight array. The information from this array  is bundled into 
576 segments covering the full central barrel. 
Each of these segments covers an area of approximately 30x50 cm$^{2}$
and  delivers one bit per beam bunch crossing depending
on whether a track has been seen within the segment. A logic trigger
unit collects the 576 bits and can set multiplicity conditions and
topological constraints. In addition, the information of the V0A and
V0C detectors is available at this level hence the required gap width 
can be defined. The output of this trigger unit is fast enough 
to reach the ALICE central trigger processor well before the time limit
for L0 decision.
   
The information of the zero degree calorimeter  can be used in the
high-level trigger HLT to identify different diffractive event classes. 
Reactions of the type $pp \rightarrow ppX$ do not carry a signal in
the zero degree calorimeters. Here, X denotes a centrally produced
diffractive state. Events of the
type $pp \rightarrow pN^{*}X$ are characterized by a signal in one of
the two calorimeters whereas events $pp \rightarrow N^{*}N^{*}X$ carry
a signal in both calorimeters. 
      
\section{Signatures of Pomeron}
\label{sec:pomeron}

The geometry of the ALICE experiment is suited for measuring a centrally
produced diffractive state with a rapidity gap on either side. Such a 
topology results from double Pomeron exchange with subsequent 
hadronization of the central state. It is expected that the
secondaries from these Pomeron-Pomeron fusion events show markedly 
different characteristics as compared to secondaries from inelastic
minimum bias events.

First, it is expected that the production cross section of glueball states
in Pomeron fusion is larger as compared to inelastic minimum bias
events. It will therefore be interesting to study the resonances produced
in the central region when two rapidity gaps are required\cite{close}. 

Second, the slope $\alpha'$ of the Pomeron trajectory is rather small:
$\alpha' \sim$ 0.25 GeV$^{-2}$ in DL fit and  $\alpha' \sim$ 0.1
GeV$^{-2}$ in vector meson production at HERA\cite{DL}. These values of
$\alpha'$ in conjunction with the small t-slope ($<$ 1 GeV$^{-2}$ ) 
of the triple Pomeron vertex indicate that the mean transverse
momentum $k_t$ in the Pomeron wave function is relatively large 
$\alpha' \sim$ 1/$k_t^2$, most probably \mbox{$k_t >$ 1 GeV}. The transverse 
momenta of secondaries produced in Pomeron-Pomeron interactions are of 
the order of this $k_t$. Thus the mean transverse momenta of secondaries 
produced in Pomeron-Pomeron fusion is expected to be larger as compared to 
inelastic minimum bias events. 

Third, the large $k_t$ described above 
corresponds to a large effective temperature. A suppression of strange 
quark production is not expected. Hence the K/$\pi$ ratio is expected to
be enhanced in Pomeron-Pomeron fusion as compared to inelastic minimum 
bias events. Similarly, the $\eta$/$\pi$ and $\eta'$/$\pi$ ratios are 
expected to be enhanced due to the hidden strangeness content and due 
to the gluon components in the Fock states of $\eta,\eta'$.

\section{Signatures of Odderon}
\label{sec:odderon}

The Odderon was first postulated in 1973 and is represented  
by color singlet exchange with negative C-parity\cite{nicolescu}. Due to its
negative C-parity, Odderon exchange can lead to differences between
particle-particle and particle-antiparticle scattering. In QCD, 
the Odderon can be a three gluon object in a symmetric color state.
Due to the third gluon involved in the exchange, a suppression by the 
coupling $\alpha_s$ is expected as compared to the two gluon Pomeron
exchange. However, finding experimental signatures of the Odderon 
exchange has so far turned out to be extremely difficult\cite{ewerz}.
A continued non-observation of Odderon signatures 
would put considerable doubt on the formulation of high energy
scattering by gluon exchange\cite{pomeron_qcd}. The best evidence 
so far for Odderon 
exchange was established as a difference between the differential
cross sections for elastic $pp$ and $p\bar{p}$ scattering 
at $\sqrt{s}$ = 53 GeV at the CERN ISR. The $pp$ cross section
displays a dip at t = -1.3 GeV$^2$ whereas the $p\bar{p}$ cross
section levels off. Such a behaviour is typical for negative
C-exchange and cannot be due to mesonic Reggeons only.   

Signatures of Odderon exchanges can be looked for in exclusive 
reactions where the Odderon (besides the Photon) is the only possible
exchange. Diffractively produced C-even states such as pseudoscalar
or tensor mesons can result from Photon-Photon, Photon-Odderon and
Odderon-Odderon exchange. Any excess measured beyond the well
understood Photon-Photon contribution would indicate an Odderon
contribution.

Diffractively produced C-odd states such as vector mesons 
$\phi, J/\psi, \Upsilon$ can result from Photon-Pomeron or 
Odderon-Pomeron exchange. Any excess beyond the Photon contribution
would be indication of Odderon exchange. 

Estimates of cross section for diffractively produced $J/\psi$ in pp
collisions at LHC energies were first given by Sch\"{a}fer et 
al\cite{schaefer}. More refined calculations by Bzdak et al result in 
a t-integrated photon contribution of $\frac{d\sigma}{dy}\mid_{y=0} \;
\sim$ 15 nb and a t-integrated Odderon contribution of   
$\frac{d\sigma}{dy}\mid_{y=0} \; \sim$ 1 nb\cite{bzdak}. 
These two numbers carry large uncertainties, the upper and lower 
limit of these numbers vary by about an order of magnitude. This cross 
section is, however, at a level where in 10$^6$ s of ALICE data taking the 
$J/\psi$ can be measured in its e$^+$e$^-$ decay channel at a level 
of 4\% statistical uncertainty. Due to the different t-dependence, 
the Photon and Odderon contribution result in different transverse
momentum distribution $p_T$ of the $J/\psi$. A careful transverse 
momentum analysis of the $J/\psi$ might therefore allow to disentangle
the Odderon contribution. 

If the diffractively produced final state is not an eigenstate of
C-parity, then interference effects between Photon-Pomeron and
Photon-Odderon amplitudes can be analyzed\cite{brodsky}. Charge asymmetry
in pion or kaon pairs is thought to be sizable\cite{haegler,ginzburg}. 
From the transverse momenta of the two particles in the pairs, 
the vectors of sum and difference
can be calculated. The sum is C-even whereas the difference is C-odd.
The opening angle $\alpha$ between sum and difference vector behaves 
as $\alpha \rightarrow \alpha + \pi$ under C-parity, hence 
a Fourier analysis of the $\alpha$-distribution will allow to
quantify the C-odd contribution.

\section{Photoproduction of heavy quarks}
\label{sec:photo}

Diffractive reactions involve scattering on small-x gluons in the
proton. The number density  of gluons at given x increases with Q$^2$, 
as described by the DGLAP evolution. Here, Q$^2$ and x denote the
kinematical parameters used in deep inelastice ep scattering. The
transverse gluon density at a given Q$^2$ increases with decreasing x as
described by the BFKL evolution equation. At some density, gluons will
overlap and hence reinteract. In this regime, the gluon density
saturates and the linear DGLAP and BFKL equation reach their range of
applicability. A saturation scale Q$_s$(x) is defined which represents 
the breakdown of the linear regime. Nonlinear effects become visible
for Q $<$ Q$_s$(x). 

Diffractive heavy quark photoproduction represents an interesting probe to
look for gluon saturation effects at LHC. The inclusive cross section
for $Q\bar{Q}$ photoproduction can be calculated within the dipole
formalism. In this approach, the photon fluctuates into a $Q\bar{Q}$
excitation which interacts with the proton as a color dipole. The
dipole cross section $\sigma$(x,r) depends on x as well as on the
transverse distance r of the $Q\bar{Q}$ pair. A study of inclusive
heavy quark photoproduction in pp collisions at LHC energy has been 
carried out\cite{goncalves1}. 
These studies arrive at differential cross sections for open charm 
photoproduction of $\frac{d\sigma}{dy}\mid_{y=0} \; \sim$ 1.3 $\mu$b 
within the collinear pQCD approach as compared to 
$\frac{d\sigma}{dy}\mid_{y=0} \; \sim$ 0.4 $\mu$b within the color 
glass condensate (CGC). The cross sections are such that open charm  
photoproduction seems measurable with good statistical significance.
The corresponding numbers for the cross section for bottom 
photoproduction are $\frac{d\sigma}{dy}\mid_{y=0} \;
\sim$ 20 nb and 10 nb, respectively.

Diffractive photoproduction is characterized by two rapidity gaps in
the final state. In the dipole formalism described above, the two
gluons of the color dipole interaction are in color singlet state. 
Diffractive heavy quark photoproduction cross sections in 
pp, pPb and PbPb collisions at LHC have been studied\cite{goncalves2}.
The cross sections for diffractive charm photoproduction are 
$\frac{d\sigma}{dy}\mid_{y=0} \; \sim $ 6 nb in pp,
$\frac{d\sigma}{dy}\mid_{y=0} \; \sim $ 9 $\mu$b in pPb and
$\frac{d\sigma}{dy}\mid_{y=0} \; \sim $ 11 mb in PbPb collisions.
The corresponding numbers for diffractive bottom photoproduction are 
$\frac{d\sigma}{dy}\mid_{y=0} \; \sim $ 0.014 nb in pp,
$\frac{d\sigma}{dy}\mid_{y=0} \; \sim $ 0.016 $\mu$b in pPb and
$\frac{d\sigma}{dy}\mid_{y=0} \; \sim $ 0.02 mb in PbPb collisions.
 
Heavy quarks with two rapidity gaps in the final state can, however,
also be produced by central exclusive production, i.e. two Pomeron
fusion. The two production mechanisms have a different t-dependence. A
careful analysis of the transverse momentum $p_T$ of the $Q\bar{Q}$
pair might therefore allow to disentangle the two contributions. 
 
\vspace{1cm}

{\bf Acknowledgments}

This work was supported in part by German BMBF under project 06HD197D.






\end{document}